\documentclass[%
reprint,
superscriptaddress,
preprintnumbers,
nofootinbib,
amsmath,amssymb,
aps,
prl,longbibliography,
]{revtex4-2}
\usepackage{graphicx} 
\usepackage{microtype}
\usepackage[T1]{fontenc}
\usepackage[normalem]{ulem}\usepackage{amsmath}
\usepackage{braket}
\usepackage{bm} 
\usepackage{comment}
\usepackage{dsfont} 
\usepackage{graphicx} 
\usepackage[dvipsnames]{xcolor}
\usepackage{hyperref}
  \hypersetup{
    colorlinks=true,
    linkcolor=BrickRed,
    filecolor=BrickRed,
    citecolor=BrickRed,
    urlcolor=RoyalBlue,
  }
\usepackage[capitalize, nameinlink]{cleveref}
\usepackage[figure, figure*]{hypcap}
\usepackage[makeroom]{cancel}
\usepackage{hyperref}
\usepackage{booktabs}

\usepackage{multirow}

\makeatletter
\let\frontmatter@footnote@produce\frontmatter@footnote@produce@footnote
\def\@bibstyle{apsrev4-2-3auth}
\makeatother

\newcommand{\bfD}{{\bm{D}}}
\newcommand{\bfJ}{{\bm{J}}}

\newcommand\be{\begin{equation}}
\newcommand\ee{\end{equation}}
\def\bea{\begin{eqnarray}}
\def\eea{\end{eqnarray}}

\newcommand{\eql}[1]{\label{eq:#1}}

\newcommand{\sfig}[2]{
\begin{center}
\includegraphics[width=#2]{#1}
\end{center}
        }

\newcommand{\Spng}[2]{
    \begin{figure}[htb]
    \sfig{./figs/#1.png}{0.8\columnwidth}
    \caption{{\small #2}}
    \label{fig:#1}
    \end{figure}
}

\newcommand\rf[1]{\ref{fig:#1}}

\begin{document}
\title{Disentangling constraints on $S_8$ from the high and low redshift CMB}
\date{\today}

\author{Joann Jones}
\email{joannjones@uchicago.edu}
\affiliation{Kavli Institute for Cosmological Physics and the Department of Astronomy \& Astrophysics, University of Chicago, Chicago, IL 60637, USA}
\author{Ualype de Andrade Uchôa}
\affiliation{Kavli Institute for Cosmological Physics and the Department of Astronomy \& Astrophysics, University of Chicago, Chicago, IL 60637, USA}
\author{Scott Dodelson}
\affiliation{Fermi National Accelerator Laboratory, P.O. Box 500, Batavia, IL 60510, USA}
\affiliation{Kavli Institute for Cosmological Physics and the Department of Astronomy \& Astrophysics, University of Chicago, Chicago, IL 60637, USA}

\begin{abstract}
    The clustering of matter in the universe is quantified by $\sigma_8$, the RMS fluctuations today in linear theory in spheres of radius $8h^{-1}$ Mpc, where $h$ parametrizes the Hubble constant. Surveys of gravitational lensing and galaxy clustering best constrain $S_8\equiv (\Omega_m/0.3)^{0.5} \sigma_8$, where $\Omega_m$ is the energy density of matter in units of the critical density. Comparing this late time measurement to the value of $S_8$ inferred from the cosmic microwave background (CMB), one can test the $\Lambda$CDM prediction for the growth of clustering from the primordial universe to now. Several analyses report hints that CMB measurements of $S_8$ are larger than those from galaxy surveys. However, CMB measurements have information from both high and low redshift due to lensing, and these tests traditionally conflate the two. Here, we present a technique to extract constraints on the cosmological parameters from each redshift component separately, enabling a high-\(z\) vs. low-\(z\) clustering comparison that was not previously possible. We marginalize over lensing information using a principal component analysis, and show that the primordial CMB from combined Planck 2018, ACT DR6, and SPT-3G D1 prefers $S_8= 0.795\pm0.022$, in agreement with the recent determination from the Dark Energy Survey (DES) of $S_8=0.789\pm0.012$. The clustering inferred from CMB lensing is larger. However, both the level of excess lensing and increase in agreement with DES depends on the prior on the sum of neutrino masses, $\sum m_\nu$. We find that removing lensing brings the CMB towards agreement with DES in any case.

    
\end{abstract}
\maketitle

{\it Introduction}---
Perhaps one of the greatest successes of the current standard cosmological model, $\Lambda$CDM, is its description of how the large-scale structure observed today emerged via gravitational collapse onto small inhomogeneities that were present in the very early universe. This claim has moved beyond a qualitative, almost obvious point, to one that can be tested with precise quantitative measurements. The level of inhomogeneity at early times, inferred from measurements of the cosmic microwave background (CMB), predicts, through $\Lambda$CDM, the precise amount of late time structure that galaxy surveys should observe today. The quantity best constrained by these late time surveys is $S_8\equiv (\Omega_m/0.3)^{0.5} \sigma_8$\footnote{We use this convention throughout for consistency and comparison to galaxy surveys. CMB lensing is best measured along $S_8^{\text{CMBL}} \equiv  (\Omega_m/0.3)^{0.25} \sigma_8$; Appendix~\ref{sec:ema} presents results in $S_8^{\text{CMBL}}$ and shows that our conclusions do not depend on the choice.}, where $\Omega_m$ is the density of matter in units of the critical density, $\sigma_8$ is the RMS linear theory density fluctuation today in spheres of radius $8h^{-1}$ Mpc, and $h\simeq 0.7$ is the dimensionless Hubble parameter $H_0 = 100h$kms$^{-1}$Mpc$^{-1}$ \cite{jainCosmologicalModelPredictions1997}.
\newcommand\zcmbl{z^{\rm CMBL}}
\newcommand\zgs{z^{\rm GS}}

The CMB and galaxy surveys roughly agree on the value of $S_8$, which may eventually be seen as a great triumph of modern cosmology. However, there is a hint of disagreement that has persisted through multiple experiments and analysis techniques for close to a decade~\cite{Pantos:2026koc}. Compared to CMB measurements, the three largest weak lensing surveys have regularly reported lower values of $S_8$~(e.g., \cite{Sugiyama:2023fzm, miyatake2023hypersuprimecamyear3, Heymans:2020gsg, DES:2026fyc}), while the final KiDS cosmic-shear analysis finds a higher value that is consistent with the CMB~\cite{Wright:2025xka}. The most recent measurement comes from the final analysis of the Dark Energy Survey (DES)~\cite{DES:2026fyc}, which reports $S_8=0.789\pm0.012$, 2.6$\sigma$ lower than the value $S_8=0.836^{+0.012}_{-0.013}$ from a CMB analysis using the same priors and free parameters. This disagreement is depicted by the green and purple contours in Fig.~\rf{fig_omegabc_S8} and the $1\sigma$ bars in Fig.~\rf{fig_s8_by_redshift}. 

\Spng{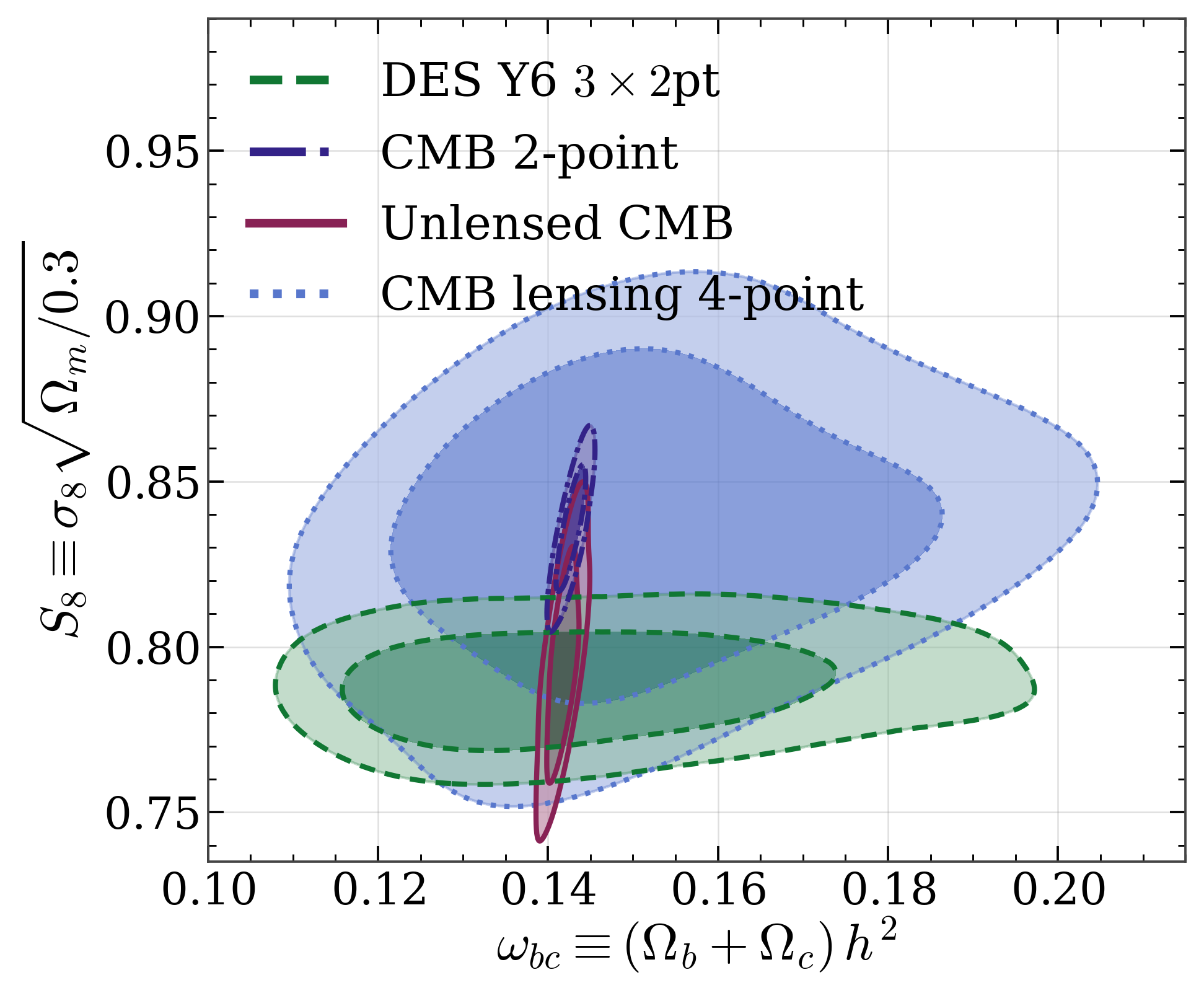}{Constraints on the physical matter density,
$\omega_{bc}\equiv(\Omega_b+\Omega_c)h^2$ and $S_8$ (68\% and 95\% credible regions). The CMB 2-pt constraint from Planck 2018, ACT DR6, and SPT-3G D1 (purple) is in mild tension with the DES Year 6 $3\times2$pt result at $0.3<\zgs<1$ (green)~\cite{DES:2026fyc}. The unlensed CMB (red), with the lensing smoothing marginalized over two principal components, measures clustering at $z_*$ and is consistent with DES. CMB lensing from the 4-pt function (blue) measures clustering at $0.7<\zcmbl<5$ and prefers the higher $S_8$ of the CMB 2-pt result. Presented with $S_8^{\text{CMBL}}$ in Appendix~\ref{sec:ema}.} 

The CMB constrains cosmological parameters mostly through the observed pattern of anisotropies at the epoch of recombination, $z_*\simeq1100$. However, the photons used to measure these anisotropies are deflected by fluctuations in the matter density along the line of sight, altering the pattern we observe, which in turn, affects the inferred posteriors on cosmological parameters. Given a set of cosmological parameters, $\Lambda$CDM predicts how strongly the primary anisotropies should be distorted \cite{blanchardGravitationalLensingEffect1987, coleGravitationalLensingFluctuations1989, lewis_weak_2006, DodelsonSchmidt2021, seljakGRAVITATIONALLENSINGEFFECT1996}. This effect is referred to as CMB lensing. 

There have been multiple reports of {\it excess} lensing in the CMB and interpretations of the implications~\cite{collaboration_planck_2017, collaboration_planck_2020, jhaveri_turning_2025, valentinoCosmologicalHintsModified2016, estebanNonstandardNeutrinoCosmology2022,renziCorneringPlanck$A_lens$2018, motloch_lensing-like_2020, calabrese_cosmic_2008,Graham:2025dqn, naredo-tueroLivingEdgeCritical2024}.  
The redshifts at which CMB photons are lensed, which we refer to as $\zcmbl$, range from $0.7<\zcmbl<5$, 
peaking at $z=2$~\cite{lewis_weak_2006, kirkCrosscorrelationGravitationalLensing2015, Madhavacheril2024_ACT_DR6_Cosmology}. This is an essential point in discussions surrounding the $S_8$ tension, which compare the amount of clustering predicted by the CMB with that observed in galaxy surveys. If some of the CMB's constraining power on $S_8$ comes from lensing, then we have been implicitly conflating high-$z$ and low-$z$ information \cite{lemosCMBConstraintsEarly2023}.

At present, it is not clear whether either of these -- the $S_8$ tension or CMB lensing excess -- is more than a statistical fluctuation. However, with the precision expected from the next generation of surveys, it will be important to identify the sharpest, most robust tests of $\Lambda$CDM. By separating early and late time information in the CMB, we put claims of excess lensing in context and explore how they impact the $S_8$ discussion. Using the high-resolution ACT and SPT-3G spectra, we show that the smoothing of the acoustic peaks via lensing {\it cannot} be mimicked by shifts in the unlensed spectra when $\Lambda$CDM parameters are varied. Our primary result suggests that the clustering inferred from $z_*$ is entirely consistent with the low-redshift ($0.3<\zgs<1$) clustering measurements from galaxy surveys (GS), specifically DES Y6 \cite{DES:2026fyc}. The present day clustering inferred from $\zcmbl$ is higher than both $z_*$ and $\zgs$, but we demonstrate that this apparent excess depends on the prior for the neutrino mass. We conclude that if there is a tension, then that tension is between high clustering at $\zcmbl$ and low clustering at both $z_*$ and $\zgs$. 

{\it CMB lensing and data}---
CMB lensing leaves two imprints on the anisotropy pattern: (i) deflections smear power between neighboring multipoles, smoothing the peaks and troughs in the standard 2-pt spectra $C_\ell^{TT,TE,EE}$ \cite{seljakGRAVITATIONALLENSINGEFFECT1996, lewis_weak_2006, metcalfGravitationalMagnificationCosmic1997, bernardeauWeakLensingStatistics1997, zaldarriagaGravitationalLensingEffect1998, seljakWeakLensingReconstruction1998} and (ii) inhomogeneities due to the gravitational potential $\Phi$ varying along different lines of sight produce off-diagonal elements in the 2-pt functions~\cite{hu_mass_2002, okamotoCMBLensingReconstruction2003}. Put another way, in the absence of lensing the standard 2-pt function $\langle a_{\ell m} a^*_{\ell'm'}\rangle$ is non-zero only when $\ell=\ell'$ and $m=m'$, and the resulting coefficient $C_\ell$ quantifies the power. This function is smoothed by lensing. In addition, gravitational lensing induces off-diagonal terms, making $\langle a_{\ell m}a^*_{\ell'm'}\rangle$ non-zero for $\ell\neq\ell'$ \cite{DodelsonSchmidt2021}. An optimal quadratic combination of those off-diagonal terms \cite{Hu:2001tn, hu_mass_2002} reconstructs the lensing potential and yields its power spectrum $C_L^{\Phi\Phi}$ directly. The smoothing effect is embedded in the measured (diagonal) 2-pt function, while the reconstruction is a 4-pt measurement; the two are not the same observable. However, they both probe the same epoch $\zcmbl$.

We are interested in separating high ($z_*$) and low ($\zcmbl$) redshift CMB constraints. This requires a clean removal of lensing effects while retaining information on the unlensed cosmological parameters. First, we infer parameter constraints from $\zcmbl$ using the 4-pt function data and running a chain on varying the $\Lambda$CDM parameters. For the 2-pt function, we remove the smoothing using a principal component analysis (PCA) \cite{Motloch:2018pjy}. We separate the lensing information in the 2-pt function from the unlensed CMB and report $S_8^{z_*}$, the late-time clustering amplitude inferred from the unlensed CMB at $z_*$, extrapolated to \(z=0\) via $\Lambda$CDM. In Appendix~\ref{sec:emc} we show that the information marginalized this way is almost entirely lensing, leaving us with a true $z_*$ constraint that has not been previously achievable.

To arrive at our constraints, we use the \textsc{CosmoSIS} cosmological parameter Bayesian inference software \cite{zuntz_cosmosis_2015}, compute the CMB and lensing power spectra with \textsc{camb} \cite{Lewis:1999bs, Howlett:2012mh}, and sample the posteriors with the importance nested sampler \textsc{nautilus} \cite{lange_nautilus_2023}.  

Throughout, we use the same combination of CMB experiments: Planck 2018~\cite{refId0}, ACT DR6~\cite{AtacamaCosmologyTelescope:2025blo}, and SPT-3G D1~\cite{Camphuis_2026}. This combination was used in the DES Year 6 analysis of galaxy clustering and lensing~\cite{DES:2026fyc}, the so-called ``$3\times2$pt'' analysis, which we use to provide a late-time constraint on clustering. We combine the Planck 2018 \textsc{plik\_lite} likelihood \cite{prince_data_2019} over $2 < \ell_{\rm TT} < 1000$ and $30 < \ell_{\rm TE,EE} < 600$; the Planck \textsc{SimAll} low-$\ell$ $EE$ likelihood \cite{collaboration_planck_2020} over $2 < \ell_{\rm EE} < 30$, which supplies the constraint on $\tau$, the ACT DR6 CMB-only (``lite'') likelihood over $600 < \ell < 7800$ \cite{AtacamaCosmologyTelescope:2025blo}, and the SPT-3G D1 \texttt{SPT3G\_D1\_TnE} lite likelihood evaluated with \textsc{candl}~\cite{balkenhol_candl_2024}, over $400 < \ell_{\rm TT} < 3000$ and $400 < \ell_{\rm TE,EE} < 4000$. For the 4-pt function, we use the combined \textsc{actplanckspt3g\_baseline} variant of the likelihood \cite{Qu:2026:JointCMBLensing, geCosmologyCMBLensing2025, Madhavacheril2024_ACT_DR6_Cosmology, Qu2024_ACT_DR6_LensingMaps, MacCrann2024_ACT_DR6_Systematics, Carron2022_PlanckPR4}. 

We use flat-$\Lambda$CDM parameters ($A_s, \Omega_m, \Omega_b, h, n_s, \sum m_\nu,\tau$) with the wide top-hat priors used in the DES Year 6 analysis \cite{DES:2026fyc}. Here, $A_s$ is the amplitude of the primordial power spectrum, $n_s$ the slope of the spectrum, $\Omega_i$ the ratio of the density of the $i^{\rm th}$ species to the critical density, $\sum m_\nu$ the sum of the neutrino masses, and $\tau$ the optical depth to reionization. In runs including only the 4-pt likelihood, we keep $\tau$ fixed at the Planck 2018 value of $0.054$, since CMB lensing is largely insensitive to reionization physics~\cite{collaboration_planck_2020}.

{\it Clustering inferred from $\zcmbl$ from the 4-pt function}--- 
Using the 4-pt information alone, we find
\be
S_8^{4\text{-pt}}= 0.836 \pm 0.033,
\ee as shown in Figs.~\rf{fig_omegabc_S8} and \rf{fig_s8_by_redshift}. This suggests that the 4-pt data are consistent with the ``full'' CMB 2-pt measurement and prefer more clustering than the DES result $S_8^{\rm GS}=0.789\pm0.012$. Our interval is wider than that of~\cite{Qu:2026:JointCMBLensing} because we adopt the DES Y6 priors, leave $\sum m_\nu$ free, and add no BAO data, but the central value remains the same (see Appendix~\ref{sec:ema}).

{\it Clustering inferred from $z_*$ from the 2-pt functions}---
The CMB 2-pt functions  $C_\ell^{TT,TE,EE}$, as discussed above, contain information from both clustering from redshift $z_*$ and clustering from $\zcmbl$ via smoothing of the peaks. In order to recover high redshift parameter constraints without the effects of lensing, we remove the smoothing via marginalization over possible lensing potentials. 

One way to attempt to remove the lensing information is to rescale it, replacing $C_L^{\Phi\Phi}$ by $A_LC_L^{\Phi\Phi}$ in the calculation of the lensed 2-pt function and marginalizing over $A_L$ \cite{calabrese_cosmic_2008}. A more general approach for separating out lensing information was presented by Motloch and Hu~\cite{motlochLensCovarianceEffects2017,motlochCMBLensSample2017,Motloch:2018pjy, motloch_lensing-like_2020}. Instead of a uniform rescaling, they decompose the lensing power spectrum into its principal components (PCs). This technique allows for precise isolation of how the temperature and polarization power spectra respond to variations in the lensing potential. Specifically, they write the lensing power spectrum, $C_L^{\Phi\Phi}$, as
\be
C_L^{\Phi\Phi} = C_{L,{\rm fid}}^{\Phi\Phi}\,\exp\left\{ \sum_{i=1}^{N
_{PC}}\, K_L^{(i)} \Theta^{(i)} \right\}
\label{eq:deform}
\ee
where $C_{L,{\rm fid}}^{\Phi\Phi}$ is a fixed fiducial lensing power spectrum, $K_L^{(i)}$ are orthonormal modes ordered by how much information they contain about power spectra peak-smoothing, and $\Theta^{(i)}$ are the coefficients of these PCs. In our chains, only the coefficients $\Theta^{(i)}$ are varied. Because $C_{L,{\rm fid}}^{\Phi\Phi}$ is fixed, the lensing that smooths the peaks no longer depends on the cosmological parameters. The parameters are constrained only through the unlensed spectra, while the $\Theta^{(i)}$ absorb the smoothing.

We compute $K_L^{(i)}$ from an approximate Fisher matrix that is calibrated to reproduce the error bars for each experiment. The resulting eigenmodes $K_L^{(i)}$ are shown in  Appendix~\ref{sec:emb}
and differ slightly from those in \cite{motloch_lensing-like_2020}, due to the inclusion of the recent ACT and SPT data. 

We take $C_{L,{\rm fid}}^{\Phi\Phi}$ to be the lensing spectrum at the posterior mean of the \emph{unlensed} CMB. We first estimate this cosmology through a marginalization over $A_L$, use that best fit as a starting fiducial for a two-PC fit, and adopt the posterior mean of that fit as the final $C_{L,{\rm fid}}^{\Phi\Phi}$. We have checked that the constraints on the $\Lambda$CDM parameters from $z_*$ do not depend on the choice of the fiducial spectrum. 

Using wide flat priors on $\Theta^{(i)}$, as described by Ref.~\cite{Motloch:2018pjy}, we find that, for the 2-pt data, only the first two components are constrained. The Fisher matrix places 98.3\% of the smoothing information in $K^{(1)}$ and $K^{(2)}$, and a five-component fit leaves the rest prior-dominated without moving $S_8$ (see Appendix~\ref{sec:emb}).

The red contour in Fig.~\rf{fig_omegabc_S8} 
displays the constraint we recover in the $\omega_{bc}$--$S_8$ plane . The 1$\sigma$ constraint, 
\be
S_8^{z_*}= 0.795\pm{0.022}\eql{s8zstar},
\ee 
shown in Fig.~\rf{fig_s8_by_redshift}, is in good agreement with the DES low redshift measurement of $S_8^{\rm GS}$. Equivalently, the amplitudes of fluctuations at recombination and at $\zgs$, extrapolated to today with $\Lambda$CDM, agree with one another. We present results in the $S_8$–$\omega_{bc}$ plane because the primary CMB tightly constrains $\omega_{bc}$, so $S_8$ constrained by probes sensitive to different epochs can be compared at a common matter density, while $S_8$ retains the constraining power of $3\times2$pt. We compare this method of removing lensing with the more common method of introducing the prefactor $A_L$ in Fig.~\rf{fig_s8_by_redshift}. The $S_8^{z_*}$ agree in the two methods, and, as noted above, with DES Y6 \cite{DES:2026fyc}.

Using a similar technique, Planck \cite{collaboration_planck_2017} pointed out that since $C_L^{\Phi\Phi}$ depends on cosmological parameters, the lensing potential used to smooth the spectrum will be at least partly determined by the $z_*$ physics that also constrains those parameters. They report that lensing accounts for only about a third of the smoothing-like shifts in $TT$, and caution that $\Lambda$CDM ``is not free to arbitrarily increase the lensing potential; it must do so through other parameters that also have non-lensing related effects''. In other words, it can recreate the smoothing by shifting the cosmological parameters so that the unlensed spectrum is smoother, and the recovered constraints would not truly represent $z_*$ physics.

We recover their result when the PC decomposition is restricted to the Planck TT data alone. But, this is no longer true using the full Planck$+$ACT$+$SPT-3G combination (see Appendix~\ref{sec:emc}). With high-resolution TT, TE, and EE measurements, changes in the \(\Lambda\)CDM parameters \emph{cannot} produce the characteristic lensing-smoothing pattern in the unlensed spectra: in our combined-data metric, the overlap is only 2-3\%. So, through marginalizing over that pattern, information about what the parameters do to the unlensed spectrum is not lost. That is, marginalizing over \(A_L\) or the lensing PCs effectively removes lensing information from the two-point parameter constraints, so the results presented here provide a clean measurement of the parameters determined by recombination-era physics.

\Spng{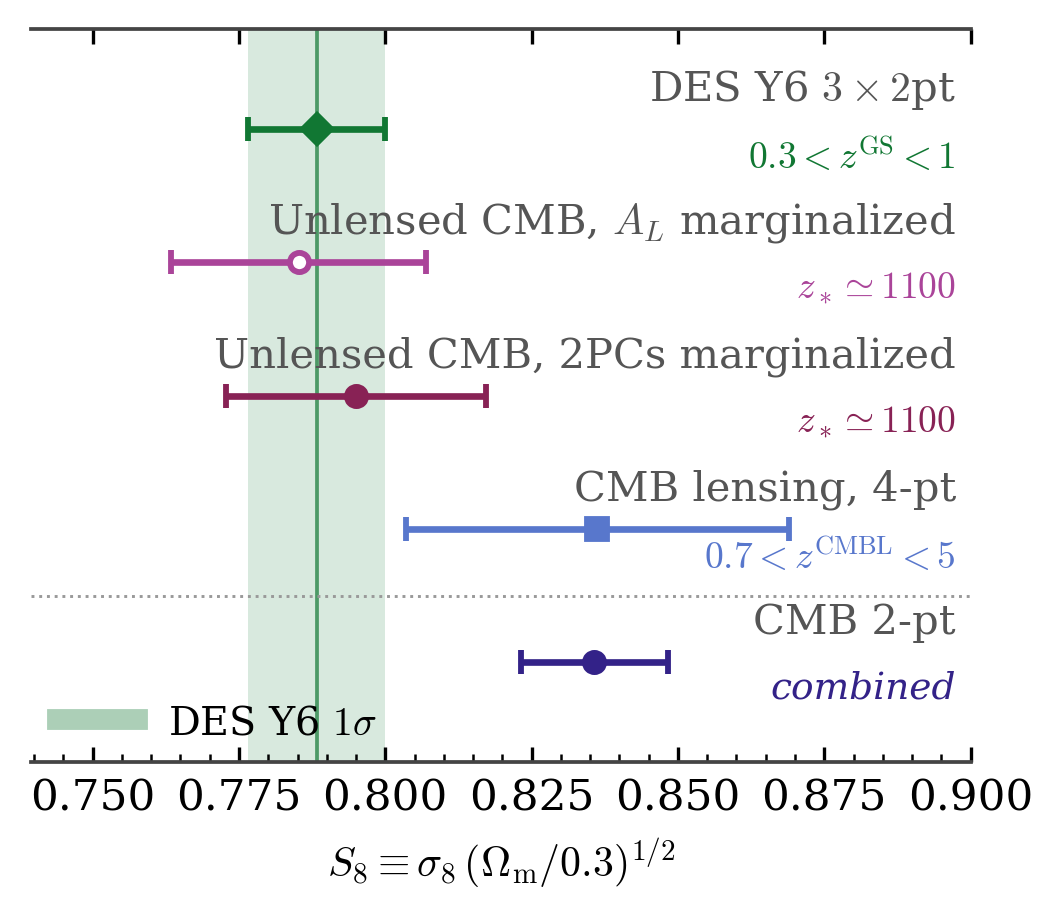}{Constraints on $S_8$ (mean and 68\% interval), labeled by the redshift each probe measures. The shaded band shows the DES Year 6 $3\times2$pt $1\sigma$ interval at $0.3<\zgs<1$. The two unlensed CMB rows measure clustering from information at $z_*$. One marginalizes the lensing smoothing over two principal components, the other marginalizes over the amplitude $A_L$. Both agree with DES. CMB lensing from the 4-pt function infers the amount of clustering from $0.7<\zcmbl<5$ and prefers a higher $S_8$, as does the CMB 2-pt result, which combines information from both epochs. See Appendix~\ref{sec:ema} for results presented in $S_8^{\text{CMBL}}.$}

{\it Neutrino mass}---
Massive neutrinos suppress the growth of structure on small scales after they become non-relativistic. Therefore, the prior on the relevant parameter ($\sum m_\nu$) can have a large impact on inferences from the CMB~\cite{LesgourguesPastor2006,Qu:2026:JointCMBLensing, green_cosmological_2024, jhaveri_turning_2025, lynch_whats_2025, cozzumboShortBlanketCosmology2025}. 
To explore this, we consider tighter priors $0.06<\sum m_\nu<X$, obtained by restricting each chain to $\sum m_\nu<X$ (Table~\ref{tab:mnu_s8}).
The second column shows that as $X$ decreases from $0.6$ to $0.1$~eV, $S_8^{z_*}$ rises by about $1\sigma$. This is because the clustering amplitude today extrapolated with $\Lambda$CDM from a fixed value at $z_*$ is {\it larger} when neutrino mass is smaller. 

The 4-pt function alone measures essentially only one combination of clustering from $\zcmbl$ so, as shown in the third column, its inference of $S_8$ is insensitive to the prior on the neutrino mass. The fourth column shows that the full CMB 2-pt function including lensing is also insensitive to the neutrino mass prior. This is for the opposite reason: with most other parameters fixed by recombination physics, lensing requires low neutrino mass, regardless of what prior is used. Finally, DES is mildly sensitive to the neutrino mass prior for the same reason as the unlensed CMB: there is less growth from the epoch of measurement ($\zgs$) to today, although the effect is smaller because $\zgs$ is much closer to today than $z_*$.


The net effect of imposing a stricter prior therefore is to slightly weaken the agreement between DES and $S_8^{z_*}$ (the difference moves from $0.3\sigma$ to $1.3\sigma$), but the difference is still much smaller than the tension with the full CMB.  

\begin{table*}
\centering
\caption{$S_8$ for flat priors $0.06<\sum m_\nu<X$~eV, obtained by restricting each chain to $\sum m_\nu<X$. The last two columns are the differences between the unlensed CMB (2 PCs marginalized) and DES Y6, and between CMB lensing and the unlensed CMB.}
\label{tab:mnu_s8}
\begin{tabular}{ccccccc}
\toprule
$X$ [eV] & Unlensed CMB & CMB lensing & CMB 2-pt & DES Y6 & Unlensed & 4-pt vs \\
 & 2 PCs marg. & 4-pt & & $3\times2$pt & vs DES & unlensed \\
\midrule
0.10 & $0.822\pm0.018$ & $0.836\pm0.032$ & $0.837\pm0.012$ & $0.796\pm0.011$ & $1.3\sigma$ & $0.4\sigma$ \\
0.30 & $0.811\pm0.018$ & $0.833\pm0.033$ & $0.836\pm0.012$ & $0.792\pm0.012$ & $0.9\sigma$ & $0.6\sigma$ \\
0.50 & $0.800\pm0.021$ & $0.834\pm0.033$ & $0.836\pm0.013$ & $0.789\pm0.012$ & $0.4\sigma$ & $0.9\sigma$ \\
0.60 & $0.795\pm0.022$ & $0.836\pm0.033$ & $0.836\pm0.013$ & $0.788\pm0.012$ & $0.3\sigma$ & $1.0\sigma$ \\
\bottomrule
\end{tabular}
\end{table*}

{\it Excess lensing}---  
When lensing is marginalized over via the inclusion of an amplitude \(A_L\) when smoothing the 2-pt function, Planck in particular exceeds the \(\Lambda\)CDM expectation (\(A_L=1\)) by over 2$\sigma$~\cite{collaboration_planck_2020}, though reanalyses have shown a shift towards \(\Lambda\)CDM~\cite{rosenbergCMBPowerSpectra2022, tristramCosmologicalParametersDerived2024}. ACT DR6, SPT-3G D1 and their combination, however, agree well with \(\Lambda\)CDM~\cite{Camphuis_2026}. Another simple test of excess lensing is to fit a free amplitude to the 4-pt function, keeping the cosmology fixed at the best-fit value obtained from the full 2-pt measurements. ACT DR6~\cite{Qu2024_ACT_DR6_LensingMaps} reports a lensing amplitude from the 4-pt function that is consistent with the Planck 2018 best-fit $\Lambda$CDM model, and the joint ACT$+$SPT$+$Planck reconstruction \cite{Qu:2026:JointCMBLensing} also finds no evidence for enhanced low-redshift structure. A similar conclusion was reported in works by SPT~\cite{omori_spt-3g_2026, geCosmologyCMBLensing2025}. 
An essential point is that these 4-pt results are presented relative to the $\Lambda$CDM best-fit model to the full CMB, which already includes lensing through the smoothing of the peaks. 

The approach described above can be applied in several ways to address this issue~(see also ~\cite{geCosmologyCMBLensing2025,omori_spt-3g_2026}). One could ask whether there is excess lensing by comparing the 4-pt posterior on $S_8$ to the unlensed $S_8^{z_*}$. As shown in Table~\ref{tab:mnu_s8}, neutrinos matter again: the stricter prior on neutrino mass brings the unlensed CMB and CMB lensing into better agreement (the difference between the two epochs falls from $1.0\sigma$ to $0.4\sigma$). As shown in Appendix~\ref{sec:ema}, this conclusion holds if the combination best measured by CMB lensing -- $\sigma_8 (\Omega_m/0.3)^{0.25}$ -- is used as a metric.

However, a better way to explore the question of excess lensing is to infer information about clustering from the posteriors on the PC coefficients $\Theta^{(i)}$. The posterior can be determined from (i) the 2-pt data, sampling all cosmological parameters to determine the unlensed spectrum and the $\Theta^{(i)}$ for $C_L^{\Phi\Phi}$; (ii) the 4-pt data, sampling only over $\Theta^{(i)}$; and (iii) by projecting the marginalized posteriors on $\Lambda$CDM parameters in the above-mentioned 2-pt chain onto the $\Theta^{(i)}$. Since we fix $C_{L, \rm fid}^{\Phi\Phi}$ at the cosmology the primary CMB prefers once its smoothing has been marginalized away, $\Theta=0$ means ``the lensing agrees with the unlensed CMB''. The 2-pt fit gives 
\begin{equation}
  \Theta_{2\text{-}\rm pt}^{(1)} = 0.97 \pm 0.20,  
  \label{eq:th2-pt}
\end{equation}
and the same deformation fit to the 4-pt bandpowers alone, with five components (Appendix~\ref{sec:emb}), gives 
\begin{equation}
\Theta^{(1)}_{\rm4\text{-}pt}=0.83\pm0.16.
 \label{eq:th4-pt}
\end{equation}
Projecting the $C_L^{\Phi\Phi}$ of 400 samples of the unlensed posterior onto the basis gives the $\Lambda$CDM prediction $\Theta^{(1)}_{\rm pred}=-0.01\pm0.36$, which the 2-pt and 4-pt exceed by $2.4\sigma$ and $2.2\sigma$, agreeing to within $0.54\sigma$. The agreement between 2-pt and 4-pt posteriors on $\Theta^{(1)}$ is robust to shifts in the prior on $\sum m_\nu$, but the predicted value from $\Lambda$CDM shifts to match. For $\sum m_\nu<0.1$~eV the prediction rises to $0.52\pm0.23$, and the 2-pt and 4-pt exceed it by only $1.4\sigma$ and $1.1\sigma$. This is due to the use of a fixed lensing spectrum when extracting these constraints.

{\it Conclusions---} 
Using the combination of Planck, ACT, and SPT-3G, the lensing smoothing of the CMB 2-pt spectra cannot be mimicked by shifts in the $\Lambda$CDM parameters. Through marginalizing over parameters that control the lensing potential, we present constraints set by recombination physics at $z_* \simeq 1100$ with minimal loss of primordial information.

Using the non-informative prior of $0.06<\sum m_\nu<0.6$~eV, the clustering amplitude today inferred from $z_*$, $S_8^{z_*}$, agrees with that inferred from $\zgs$ by DES, while the full CMB 2-pt, which combines $z_*$ and $\zcmbl$ information, differs from DES by over $2\sigma$. Additionally, the 2-pt smoothing and the 4-pt reconstruction agree and both exceed the lensing predicted from $z_*$ by more than $2\sigma$.

The framing of these differences is dependent on the choice of prior for $\sum m_\nu$. The clustering inferred from $\zcmbl$ is insensitive to the prior on $\sum m_\nu$, but the shift in the posteriors of the unlensed spectra predicts more lensing as $\sum m_\nu$ is forced lower. Restricting to $\sum m_\nu<0.1$~eV raises $S_8^{z_*}$, 
bringing it into agreement with the inference from the 4-pt function and nearly eliminating any evidence for excess lensing. But, this restriction increases the difference in $S_8$ between $z_*$ and $\zgs$ to $1.3\sigma$, which is still well below the $2.5$--$2.8\sigma$ between the full CMB 2-pt and DES. 

As new surveys continue to sharpen the low-redshift measurement, separating the CMB's information by epoch will be necessary to tell whether any remaining discrepancy lies between early and late times or among the late-time probes themselves.

{\it Acknowledgments}---
We are very grateful to Wayne Hu and Bhuvnesh Jain for fruitful discussions and to David Sanchez Cid for his generous help with running chains. We also thank Gabriel Lynch, Chihway Chang, Frank Qu, and Joel Meyers for their valuable insights, comments, and suggestions. 
J.J. acknowledges the Jump Trading Fellowship Program for partial support during the completion of this work. J.J. thanks the LSST-DA Data Science Fellowship Program, which is funded by LSST-DA, the Brinson Foundation, the WoodNext Foundation, the Research Corporation for Science Advancement Foundation, and Kevin Wells; her participation in the program has benefited this work. U.U. acknowledges the Jeff Metcalf Grant and the Quad Undergraduate Research Grant for supporting him. This work was supported by FermiForward Discovery Group, LLC, under Contract No. 89243024CSC000002 with the U.S. Department of Energy, Office of Science, Office of High Energy Physics. SD thanks the Aspen Center for Physics for its hospitality during the 2026 summer program. This work
was performed in part at the Aspen Center for Physics, which is supported by National
Science Foundation grant PHY-2210452.

This work was completed in part with resources provided by the University of Chicago’s Research Computing Center. The authors also acknowledge the Bayesian Inference software \href{https://cosmosis.readthedocs.io/en/latest/}{\textsc{CosmoSIS}} \cite{zuntz_cosmosis_2015}.

A portion of the code used in this analysis was written with assistance from the large language model (LLM), Claude Opus 5 (Anthropic). All LLM-generated code was verified and extensively tested by the authors, and no LLMs were used for result interpretation.

\appendix
\setcounter{secnumdepth}{2}

\section{Comparison to published 4-pt analyses}
\label{sec:ema}

We note that our results from the 4-pt function are slightly different from those in \cite{Qu:2026:JointCMBLensing}. Here, we describe the different analysis choices and the reasoning behind each.

A 4-pt-only posterior constrains essentially one combination of $\sigma_8$, $H_0$, and $\Omega_m$~\cite{Madhavacheril2024_ACT_DR6_Cosmology}, and the direction it best constrains is $S_8^{\rm{CMBL}}\equiv\sigma_8(\Omega_m/0.3)^{0.25}$. Using the 4-pt posterior presented here, we find that the fractional width of the constraint in $\sigma_8(\Omega_m/0.3)^{p}$ is minimized at $p=0.25$, where we measure $0.814\pm0.012$, $1.5\%$ error against $3.9\%$ when we define $p=0.5$, closely resembling what was found in  \cite{Qu:2026:JointCMBLensing,omori_spt-3g_2026, Carron2022_PlanckPR4, Qu2024_ACT_DR6_LensingMaps}. Much of the width of the blue contour in Fig.~\rf{fig_omegabc_S8} is therefore due to projection rather than measurement. But, there is no single exponent \(p\) at which CMB lensing and galaxy surveys are reported along their respective best-measured directions, and the inferred level of disagreement between them depends on which one is chosen. 

For completeness, Fig.~\rf{fig_omegabc_S8cmbl} and Table~\ref{tab:exponent_values} show our results in $S_8^{\rm CMBL}$. The CMB 2-pt is tightest at an exponent similar to the 4-pt, $p=0.28$, while DES Y6 and the unlensed CMB are tightest at $p\simeq0.6$. DES has a $1.3\%$ error at $p=0.60$ against $2.8\%$ at $p=0.25$, and the unlensed CMB $2.7\%$ at $p=0.63$ against $3.5\%$ at $p=0.25$. In $S_8^{\rm CMBL}$, CMB lensing exceeds the unlensed CMB by $1.5\sigma$, against $1.0\sigma$ in $S_8$. In this case as well, when the neutrino mass prior is tightened (to 0.1 eV), the unlensed inference on $S_8^{\rm CMBL}$ moves up significantly, in complete accord with the 4-pt inference. That is, there is no excess lensing if the neutrino mass prior is very tight. 



\Spng{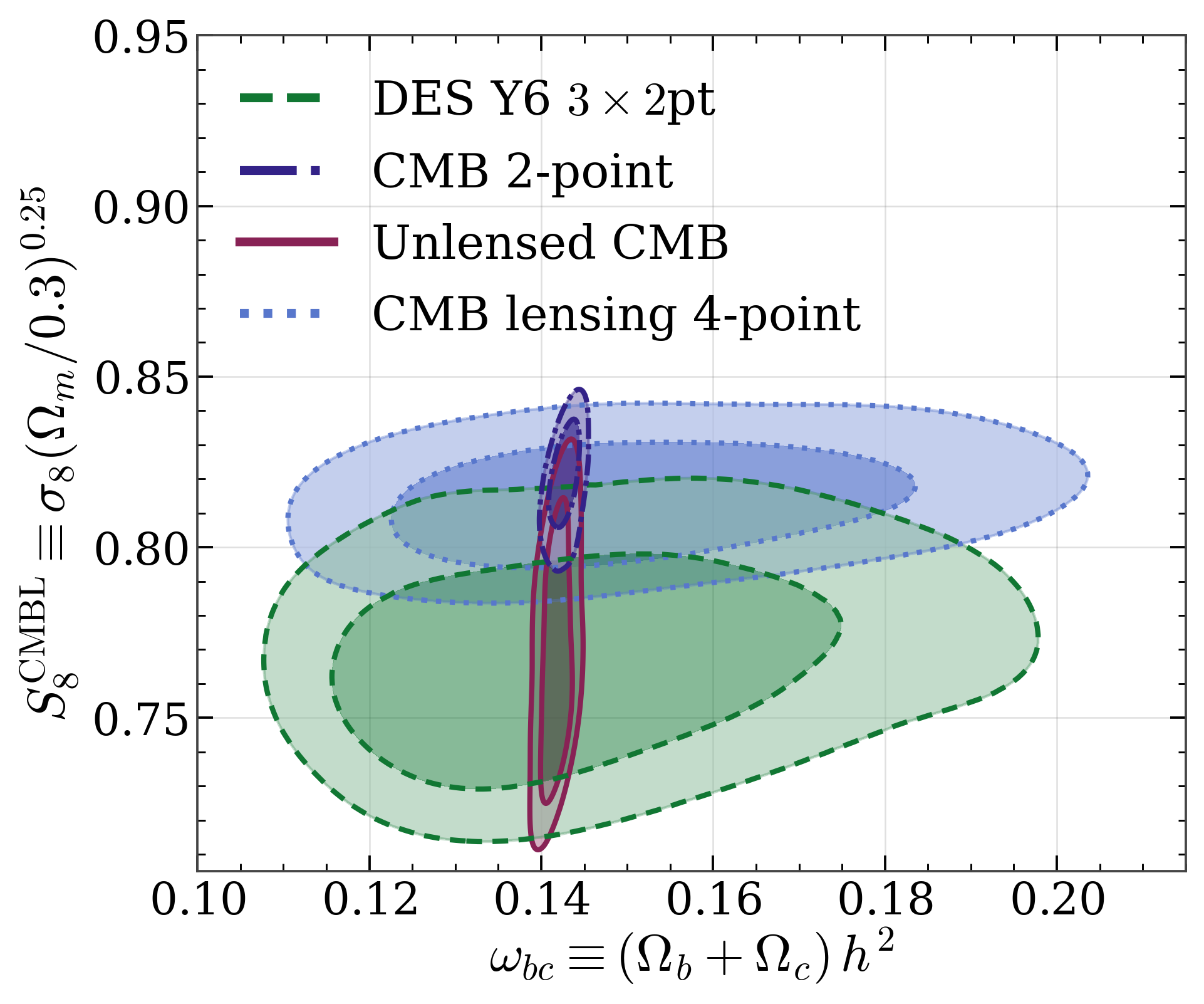}{Same as Fig.~\rf{fig_omegabc_S8}, but with $S_8^{\rm CMBL}\equiv\sigma_8(\Omega_m/0.3)^{0.25}$ on the vertical axis, highlighting the constraining power of CMB lensing (blue), which constrains this combination nearly independently of $\Omega_m$.}

Another difference is our choice to not include baryon acoustic oscillations (BAO), a dataset typically used to break the $\sigma_8$--$\Omega_m$ degeneracy \cite{Madhavacheril2024_ACT_DR6_Cosmology}. We deliberately chose not to do this since we are primarily interested in separating low and high-redshift CMB information, and BAO would add in extra information from low-redshift.

\begin{table}
\centering
\caption{Constraints in $S_8$ and $S_8^{\rm CMBL}$ with $0.06<\sum m_\nu<0.6$~eV.}
\label{tab:exponent_values}
\begin{tabular}{lcc}
\toprule
 & $S_8$ & $S_8^{\rm CMBL}$ \\
\midrule
Unlensed CMB & $0.795\pm0.022$ & $0.768\pm0.027$ \\
CMB lensing 4-pt & $0.836\pm0.033$ & $0.814\pm0.012$ \\
CMB 2-pt & $0.836\pm0.013$ & $0.820\pm0.011$ \\
DES Y6 $3\times2$pt & $0.788\pm0.012$ & $0.767\pm0.022$ \\
\bottomrule
\end{tabular}
\end{table}

These choices widen the credible interval on \(S_8\), but do not move the central result. 
Our 4-pt analysis reproduces the joint ACT$+$SPT$+$Planck constraint of $S_8^{\rm CMBL} = 0.825^{+0.015}_{-0.013}$~\cite{Qu:2026:JointCMBLensing} within $0.1\sigma$ 
if $\sum m_\nu$ is restricted to the minimal range assumed by those analyses.

\section{Number of PCs used for marginalization}
\label{sec:emb}

The components $K_L^{(i)}$ are eigenmodes of a Fisher matrix for the peak smoothing, shown in Fig.~\rf{fig_pc_basis}, meaning they are ordered by how well the 2-pt data constrain them. The first component carries 91.6\% of the information on $C_L^{\Phi\Phi}$, the first two 98.3\%, and the first five 99.9\%. We checked this against the data by repeating the 2-pt analysis with five components and find no change in the cosmological parameters. Beyond the second PC, the coefficients are not constrained: the priors are less than twice as wide as the posteriors, compared with 3.7--18.5 for the same five coefficients in the 4-pt fit. $\Theta^{(1)}$ broadens by a factor of 3.5, to $0.36\pm0.71$, so the 2-pt no longer constrains $\Theta^{(1)}$ when 5 PCs are used. The posterior means of the cosmological parameters do not move. We find $S_8^{z_*}=0.793\pm0.023$ with five components against $0.795\pm0.022$ with two, and $\sum m_\nu=0.350\pm0.152$\,eV against $0.352\pm0.151$\,eV. This is expected since $C_{L,{\rm fid}}^{\Phi\Phi}$ is fixed, so the cosmological parameters receive no lensing information for any number of components. We therefore marginalize over two PCs for our 2-pt analysis.

The 4-pt reconstruction weights multipoles differently, so an ordering that is efficient for the 2-pt is not necessarily well suited for the 4-pt. This raises the concern that a basis truncated too early could absorb the part of the signal it cannot represent by displacing the components it does retain. Here, we test whether two components suffice by measuring what a truncated analysis would report if the truth were known.

We draw samples from a chain in which the cosmological parameters vary against the 4-pt likelihood alone. Priors are flat over the sampled volume, so $\chi^2 = -2(\ln\mathcal{L}_{\rm post} - \ln\mathcal{L}_{\rm prior})$ is known for every sample without further computation, and test cases can be selected directly by their misfit to the lensing band powers. The best fitting sample has $\chi^2 = 34.7$ and the 99.9\% credible region extends to $\Delta\chi^2 = 14.9$, spanning $S_8 = 0.72$–$0.94$. We use six cosmologies for testing: the best fitting sample, three at $\Delta\chi^2 \simeq 4$, 9 and 14, and two extremes in $S_8$. 

\Spng{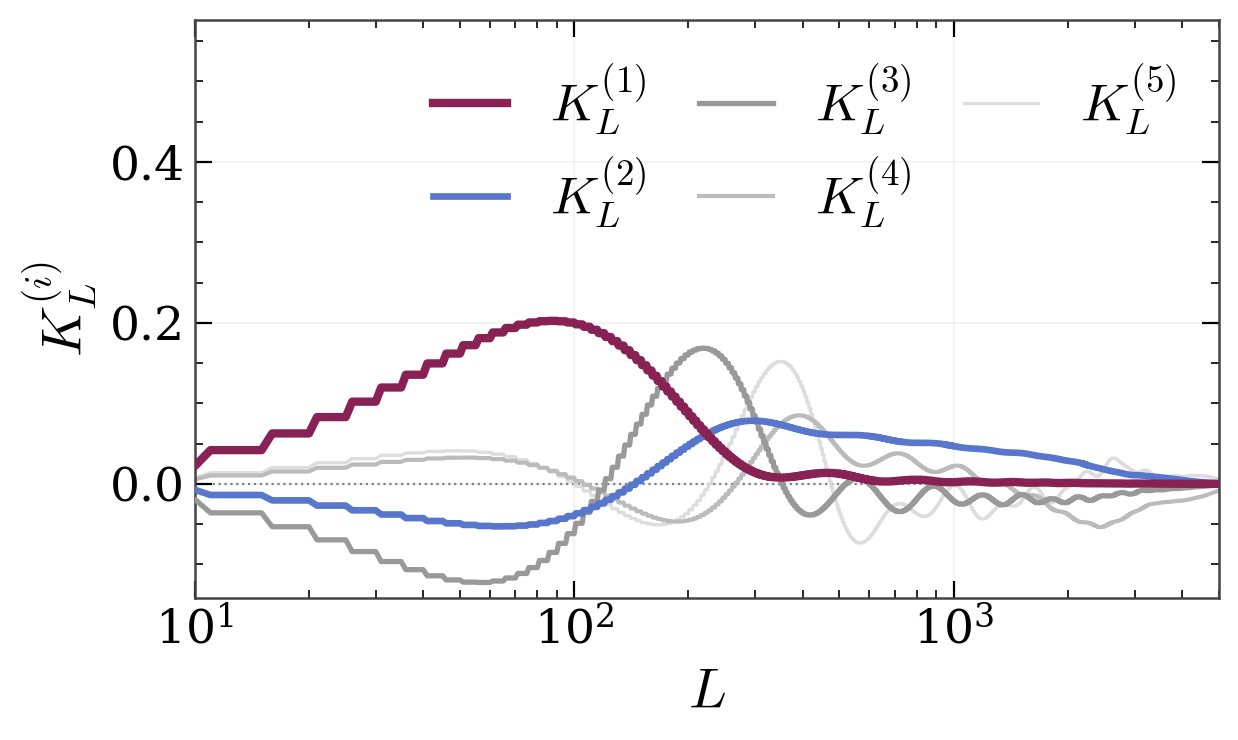}{The first five lensing PCs $K_L^{(i)}$ of
Eq.~\eqref{eq:deform}, eigenmodes of the peak-smoothing Fisher matrix. $K^{(1)}$ (red) and $K^{(2)}$ (blue) are the two marginalized over in this work. $K^{(3)}$--$K^{(5)}$ (grey) are additionally freed in the five-component fits used for the 4-pt analysis.} 

For each cosmology, we compute $C_L^{\Phi\Phi}$ exactly using \textsc{camb}. Then, we form the logarithmic deformation of the fiducial spectrum, $d(L) = \ln[C_L^{\Phi\Phi}/C_{L,\rm fid}^{\Phi\Phi}]$, and project it onto the PC basis. We then fit $N$ components to that cosmology's binned spectrum weighted by the full Planck$+$ACT$+$SPT-3G covariance. The difference $\hat\Theta^{(1)}(N) - \Theta^{(1)}$ is the bias an $N$-component analysis would incur if that cosmology sample were the truth.

We find that the truncation bias is large when two components are retained, reaching $-2.5\sigma$ in $\Theta^{(1)}$ and $-2.0\sigma$ in $\Theta^{(2)}$, but falls steeply with mode count. At $N=5$ it does not exceed $0.28\sigma$ in $\Theta^{(1)}$ or $0.18\sigma$ in $\Theta^{(2)}$. On $\Theta^{(1)}_{\rm 4-pt}=0.831\pm0.157$, this corresponds to a shift of at most $0.044$. Both coefficients remain at that level through $N=9$. Given that the stability is established across five consecutive mode counts, we retain five PCs when constraining $\Theta^{(1)}$ from the 4-pt function.

\section{Lensing or primordial information}
\label{sec:emc}

Figure~\rf{fig_theta_response} shows the 2-pt response, $\partial D_\ell^{X}/\partial\Theta^{(1)}$. Marginalizing over the PC coefficients $\Theta^{(i)}$ is equivalent to removing some modes from the 2-pt spectra. Those modes (linear combinations of $C_l^{XX}$) contain information about the cosmological parameters. As a cartoon example, imagine that only $l=3000$ is lensed. Then the principal component vector would simply be a 1 at $l=3000$ and 0 everywhere else. Marginalizing over the corresponding $\Theta$ would therefore remove any information that might be obtained from that multipole. This begs the question: when lensing is removed, how much primary information is also removed? 

\Spng{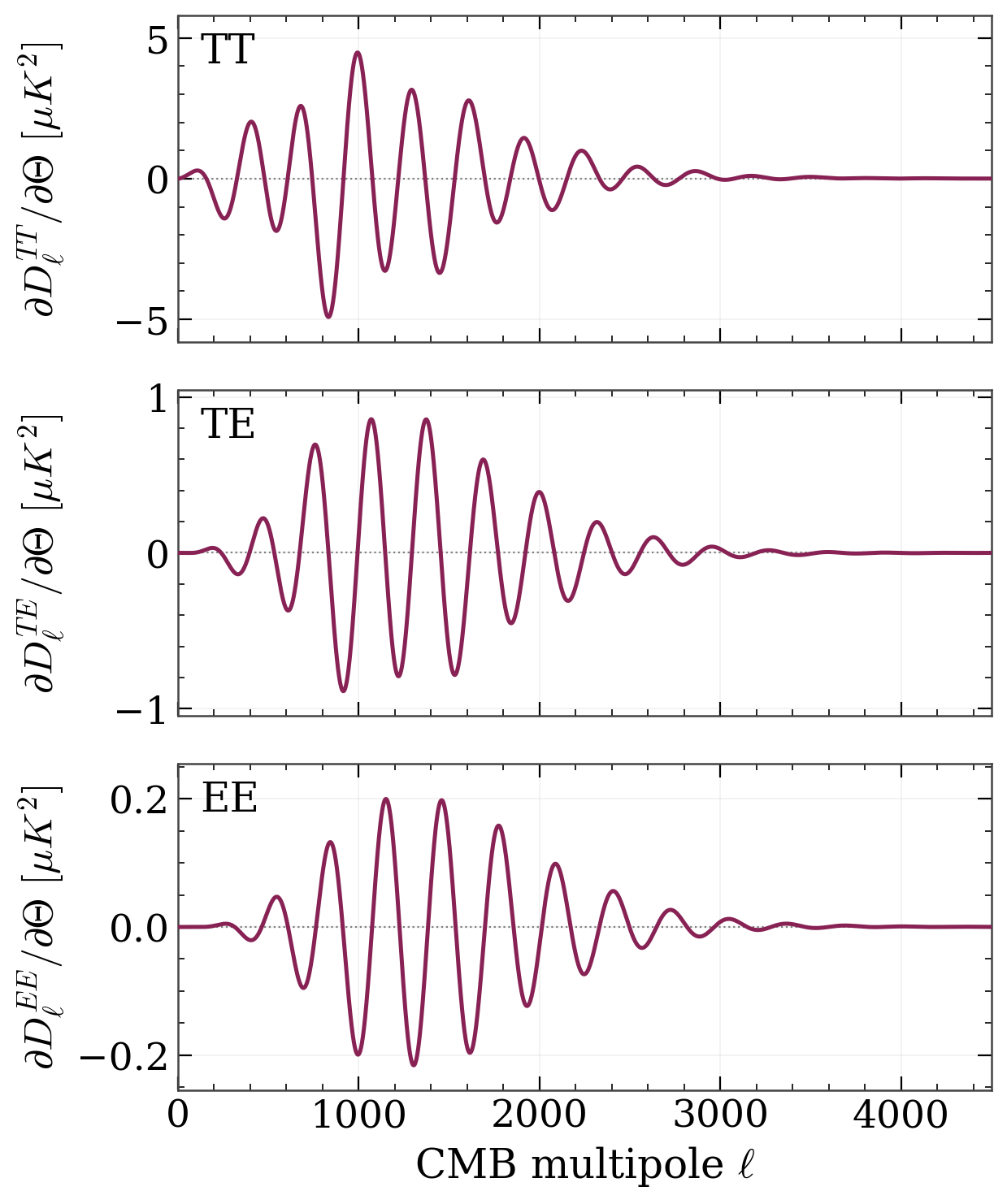}{Response of the lensed $TT$, $TE$ and $EE$ spectra to a change in $\Theta^{(1)}$ at fixed cosmology, $\partial D_\ell^X/\partial\Theta^{(1)}$, scaled to the $1\sigma$ width of the 2-pt posterior. The oscillation follows the curvature of the acoustic peaks, the signature of smoothing.}

To isolate the primary information, we linearize about the fiducial cosmology and use the Fisher matrix of the same bandpowers. Let $\bfJ_p=\partial\bfD/\partial p$ be the response of the binned spectra lensing by $C^{\Phi\Phi}_{L,{\rm fid}}$ to shifts in the seven parameters, 
and $\bfJ_\Theta=\partial\bfD/\partial\Theta^{(1)}$ the binned version of the response in Fig.~\rf{fig_theta_response}. The parameter shift that best reproduces a unit change in $\Theta^{(1)}$ is
\be
\delta p=F_{pp}^{-1}\bfJ_p^TC^{-1}\bfJ_\Theta,
\ee
with
\be
F_{pp}=\bfJ_p^TC^{-1}\bfJ_p,
\ee
and the fraction of the $\Theta^{(1)}$ signal it fails to capture is
\be
f=\frac{(\bfJ_\Theta-\bfJ_p\delta p)^TC^{-1}(\bfJ_\Theta-\bfJ_p\delta p)}{\bfJ_\Theta^TC^{-1}\bfJ_\Theta}.
\ee
For this work, we find $f=0.97$.  While the primary parameters move with $\Theta^{(1)}$ and therefore shift when it is marginalized, no combination of them reproduces the pattern of modes removed. This means that the information we marginalize over is nearly entirely isolated to lensing. If we wish to recover constraints from $z_*$ alone, this method provides that with little information lost due to marginalization. If we consider Planck alone, we find $f\sim0.3$, meaning that a shift in the cosmological parameters is able to largely replicate the effects from shifts in $\Theta^{(1)}$. This is what Ref.~\cite{collaboration_planck_2017} reported, as discussed in the main text above. With the addition of ACT and SPT, changes in the unlensed spectrum due to shifts in the $\Lambda$CDM parameters are no longer able to mimic the smoothing due to lensing. 


The agreement between the 2- and 4-pt provides another check of the information marginalized over. The 4-pt function should not absorb a non-lensing feature of the temperature spectrum, so there is no reason for the 2-pt $\Theta^{(1)}$ to land within $0.54\sigma$ of it unless the information being marginalized is primarily lensing.

\bibliography{ref}
\end{document}